\documentclass{IEEEcsmag}
\IEEEoverridecommandlockouts
\usepackage{amsmath,amsfonts}
\usepackage{subfigure}
\usepackage{graphicx}
\usepackage{adjustbox}
\usepackage{comment}
\usepackage{soul}
\usepackage{textcomp}
\usepackage{stfloats}
\usepackage{url}
\usepackage{verbatim}
\let\labelindent\relax
\usepackage{enumitem}
\usepackage{xcolor}
\definecolor{IEEEblue}{RGB}{0,102,153}
\newcommand{\authorbio}[2]{%
\vspace{0.35cm}
\noindent{\bfseries\textcolor{IEEEblue}{\MakeUppercase{#1}}}~#2\par}
\usepackage{ragged2e}
\usepackage{cite}
\usepackage{float}
\usepackage{tcolorbox}

\usepackage[colorlinks,urlcolor=blue,linkcolor=blue,citecolor=blue]{hyperref}
\expandafter\def\expandafter\UrlBreaks\expandafter{\UrlBreaks\do\/\do\*\do\-\do\~\do\'\do\"\do\-}
\usepackage{upmath}

\graphicspath{{./images/}}
\jmonth{May}
\jname{IEEE Internet Computing}
\jtitle{Orchestrating Serverless Applications in the
Edge-Cloud-Space Continuum}
\pubyear{2026}

\begin{document}
\sptitle{IEEE Internet Computing} 
\title{Orchestrating Serverless Applications in the Edge-Cloud-Space Continuum: What Breaks and What's Next?}
\author{Hadi Tabatabaee~Malazi~$^{\ddagger}$}
\affil{School of Computer Science, University College Dublin, Ireland}
\author{Reza Farahani~$^{\ddagger\ast}$}
\affil{Distributed Systems Group (DSG), TU Wien, Austria}
\author{Nitinder Mohan}
\affil{Faculty of Electrical Engineering, Mathematics, and Computer Science, TU Delft, Netherlands}
\author{Schahram Dustdar}
\affil{Distributed Systems Group (DSG), TU Wien, Austria and ICREA Barcelona, Spain\\[4pt]
\small $^{\ddagger}$ Hadi Tabatabaee Malazi and Reza Farahani contributed equally to this work.\\[4pt]
\small $^{\ast}$ Corresponding author: r.farahani@dsg.tuwien.ac.at
}
\markboth{IEEE Internet Computing}{IEEE Internet Computing}
\begin{abstract} \justifying
Serverless computing has matured into an effective execution model for edge-cloud environments, enabling function-level decomposition, demand-driven scaling, and workflow execution across stable, well-provisioned infrastructure. This success motivates extending it to the edge-cloud-space continuum, where Low Earth Orbit (LEO) constellations are increasingly explored as distributed compute substrates. However, existing serverless orchestration is not directly applicable in this setting, where LEO systems impose time-varying contact graphs, intermittent link availability, and strict feasibility constraints on energy, memory, communication, and operational cost. This article identifies ten broken assumptions in existing serverless orchestration and organizes them into three core challenges: spatiotemporal execution over dynamic graphs, constraint-aware function placement and scaling, and correctness and progress under decentralized and delayed state. It then proposes an architecture that enables robust and efficient serverless execution across the continuum, grounded in these challenges and demonstrated through a representative flood-response use case.
\end{abstract}
\maketitle
\section{From Stable Clouds to Dynamic Orbits: Why Serverless Orchestration Breaks}
\label{sec:sec1}
Serverless computing has become a dominant execution model in edge-cloud environments by abstracting resource management and enabling event-driven, function-level execution. In this context, the orchestrator determines where and when functions execute, how they scale, and how to access intermediate states through managed services. These decisions implicitly rely on stable connectivity, continuous resource availability, and timely global control. 
These assumptions break down when execution moves beyond terrestrial edge/cloud infrastructure to include Low Earth Orbit (LEO) satellites~\cite{wang2023satellite}, forming an edge-cloud-space (ECS) continuum. 
In this setting, execution is no longer purely resource-driven but is constrained by time-varying connectivity, delayed observations, and restrictions on where intermediate states can be stored and moved. The following flood response scenario makes these differences explicit.

\begin{figure*}[!t]
    \centering
    \includegraphics[width=0.97\linewidth]{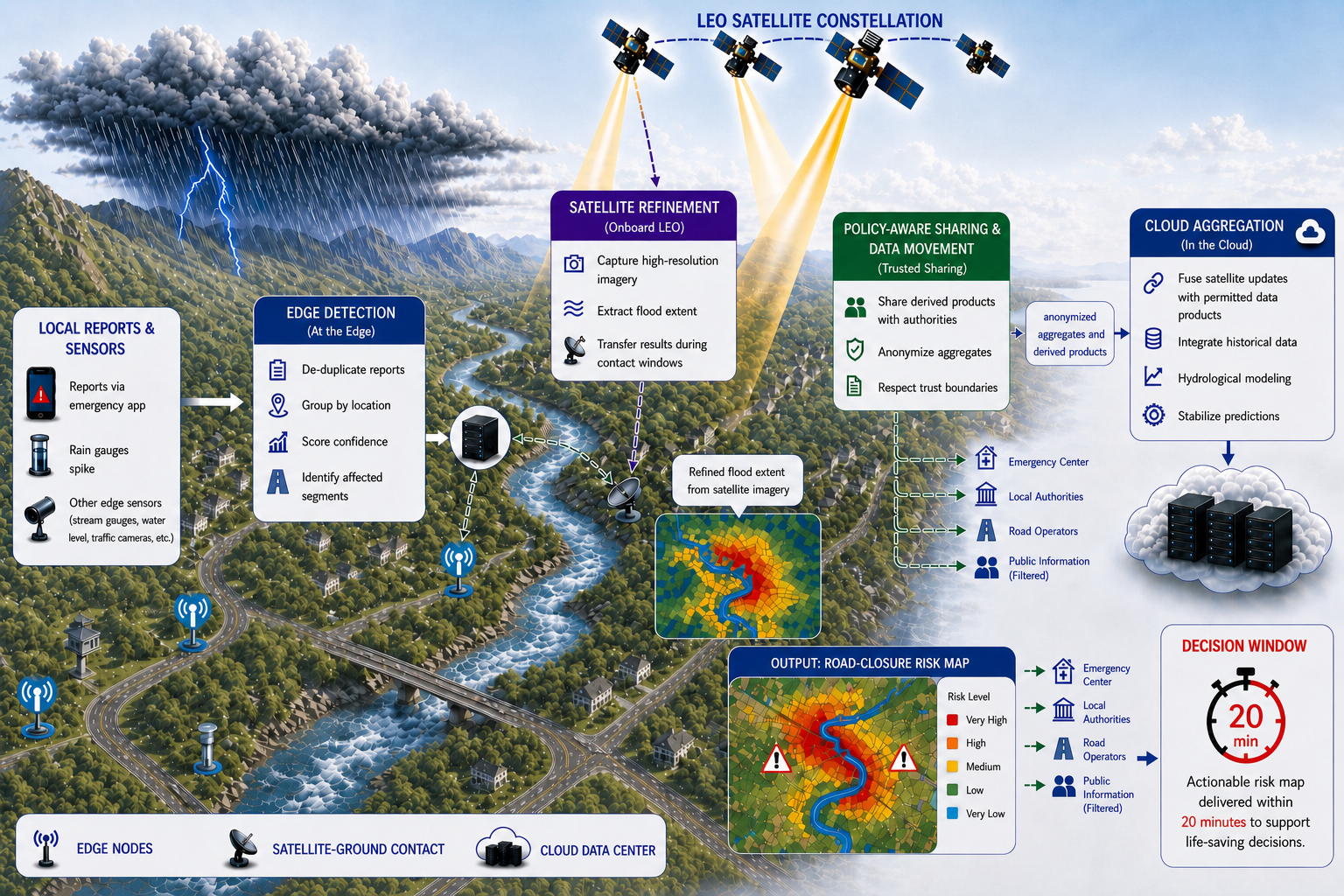}
    \caption{Serverless flood detection and response workflow across the edge-cloud-space continuum.}
    \label{fig:my_label}
\end{figure*}

Consider a severe storm hitting a rural river basin overnight. At 06:00, hundreds of residents use an emergency application to report roads being overtopped, while rain gauges show a sharp rise. Edge-side functions remove duplicate reports, group them by location, and generate a flood alert that identifies likely affected road segments and assigns an initial confidence score to the alert. The system must produce a road-closure risk map for downstream towns within a strict 20-minute decision window. The initial risk map can be generated from local reports and sensors, but an accurate assessment of flood extent and road-closure risk requires satellite imagery. This data is not continuously available. It becomes accessible only when a suitable LEO satellite has an observation or communication opportunity. The workflow thus produces an initial map and refines it as new observations become available. Cloud services fuse these updates with historical data to stabilize predictions. At the same time, governance constraints require that raw reports remain within the local domain, while only anonymized aggregates and derived products can be shared. This workflow highlights that correctness depends not only on processing data, but also on aligning computation with when data becomes available and where that data is permitted to move. Executing this workflow requires more than reactive placement and scaling. The orchestrator must transform bursty inputs into a validated trigger, schedule a multi-stage workflow under a strict deadline, and support progressive refinement as new data arrives. This introduces three tightly coupled requirements:
\vspace{-.2cm}
\begin{enumerate}[leftmargin=*, itemsep=1pt, topsep=2pt]
\item \textbf{Time-dependent execution.} Function placement and scheduling must align with communication opportunities defined by satellite contact windows and evolving paths.
\item \textbf{State-dependent execution.} The workflow relies on an evolving intermediate product, the operational flood state, whose placement and movement determine whether downstream stages can progress without repeated multi-hop transfers.
\item \textbf{Policy-constrained execution.} Trust boundaries restrict where functions can execute and how data can move, coupling orchestration decisions with governance requirements.
\end{enumerate}
\vspace{-.1cm}

In conventional cloud and edge settings, serverless platforms react to events, place functions based on available compute, and scale replicas to meet demand. The network is assumed to be continuously reachable, application state is externalized to backend services, and a centralized control plane adapts execution based on global telemetry. In such a flood response scenario, none of these assumptions holds. Connectivity is intermittent, and the topology evolves over time. Resource availability is bounded and heterogeneous, particularly in the space tier. Moreover, control decisions must often be made from stale and incomplete views of the system. Hence, correctness depends not only on where functions execute, but also on when they execute and how state is propagated across the system.
\begin{figure}[!t]
\vspace{-.2cm}
\centering
\begin{tcolorbox}[colback=orange!15,colframe=black!40!black,
        boxrule=0.8pt,arc=3pt,left=6pt,right=6pt,top=3pt,bottom=3pt]
\textbf{Takeaway.} The ECS continuum changes the nature of serverless orchestration. It introduces temporal constraints, state locality challenges, and policy coupling that invalidate the assumptions underlying existing systems.
\end{tcolorbox}
\vspace{-15pt}
\end{figure}

This article identifies ten broken assumptions in existing serverless orchestration, organizes them into concrete system-level challenges, and designs an architecture for robust and efficient execution across the ECS continuum, demonstrated through a representative flood-response use case.

\vspace{-.3cm}
\section{Broken Assumptions in Serverless Orchestration}
\label{sec:sec2}
Conventional orchestration methods are effective in cloud and terrestrial edge settings because their core assumptions, i.e., continuous reachability, stable topology, low-latency state access, elastic scaling, and centralized control, largely hold. In the ECS continuum, these assumptions break systematically due to orbital motion, intermittent connectivity, and hard resource constraints~\cite{denby2020orbital}. As a result, orchestration strategies that are valid in the cloud become infeasible or suboptimal when applied across the ECS continuum. This section identifies ten broken assumptions (\emph{BA1--BA10}) and maps them onto three core orchestration challenges (\emph{C1--C3}). These challenges define the core responsibilities of an ECS continuum-aware orchestrator and directly motivate the architecture presented in the next section.
\vspace{-.4cm}
\subsection{C1: Spatiotemporal Execution over Dynamic Graphs}
In the ECS continuum, execution feasibility is not a static property of nodes but a time-dependent one, determined by communication opportunities~\cite{bhosale2024krios}, end-to-end paths~\cite{basak2025leocraft}, and execution bindings. Orchestration must therefore ensure both computation and data exchange remain feasible over the entire execution horizon.
\vspace{-.4cm}
\subsubsection{\textbf{BA1: Instantaneous reachability implies execution feasibility.}} 
Serverless orchestration assumes that reachability at invocation time is sufficient to admit, execute, and complete a function. 
\begin{itemize}[leftmargin=*, itemsep=1pt, topsep=2pt]
 \item \textbf{Breakage.} Reachability is bounded by short contact windows~\cite{bhosale2024krios}; nodes reachable at invocation may be unreachable before execution, state transfer, or result delivery completes~\cite{10818223}.
 \item \textbf{Implication.} Execution must be \emph{temporally feasible}, aligning placement and data exchange with predicted contact windows across workflow stages. 
\end{itemize}
\vspace{-.2cm}
\subsubsection{\textbf{BA2: Connectivity implies path feasibility.}} 
Existing systems implicitly assume that network connectivity guarantees usable end-to-end communication for state access and function chaining.
\begin{itemize}[leftmargin=*, itemsep=1pt, topsep=2pt]
 \item \textbf{Breakage.} In space-ground networks, connectivity does not guarantee feasible communication~\cite{11036255}. End-to-end paths change over time, and the properties of different link types (e.g., RF vs. FSO) introduce variability in latency, throughput, and reliability~\cite{wu2025sate}. Multi-hop paths may therefore fail to meet the required data transfer deadlines~\cite{10818223}.
 \item \textbf{Implication.} The orchestrator must evaluate \emph{time-varying path feasibility}, including whether required data transfers can complete within service-level objective (SLO) constraints. 
\end{itemize}
\vspace{-.2cm}
\subsubsection{\textbf{BA3: Execution binding is stable.}} 
Serverless systems assume that once a function is placed, its execution context (node, resources, and data dependencies) remains stable during execution.
\begin{itemize}[leftmargin=*, itemsep=1pt, topsep=2pt]
 \item \textbf{Breakage.} Execution and communication bindings change as LEO satellites move and handovers occur~\cite{pfandzelter2024komet}, while compute, memory, and accelerator availability fluctuate under concurrent workloads and onboard constraints~\cite{10818223}. 
 \item \textbf{Implication.} Execution must be \emph{time-indexed and re-bindable}, supporting dynamic reassignment and continuity under changing conditions.
\end{itemize}

These broken assumptions jointly indicate that execution feasibility must be enforced over time, not at a single decision point. This leads to the first research question (\textbf{RQ1}): ``How should an orchestrator place and schedule workflow stages over a time-varying space-ground graph such that computation and data exchange remain feasible throughout execution under SLO constraints?''
Addressing this requires the orchestrator to enforce spatiotemporal feasibility as a first-class constraint, leading to the following design implications:
\begin{itemize}[leftmargin=*, itemsep=1pt, topsep=2pt]
\item \textbf{Contact window-aware scheduling:} Align execution and data transfer with predicted communication windows.
\item \textbf{Time-indexed planning:} Make placement decisions over execution horizons rather than single time points.
\item \textbf{Path-aware feasibility analysis:} Evaluate end-to-end data movement feasibility over dynamic multi-hop paths.
\item \textbf{Re-bindable execution:} Support migration and reassignment under changing associations and resource conditions.
\end{itemize}
\vspace{-.2cm}
\subsection{C2: Constraint-Aware Function Placement and Scaling} In the ECS continuum, execution is not governed by elastic resource availability but by hard feasibility constraints, including energy budget, onboard thermal headroom~\cite{xing2024deciphering}, memory capacity~\cite{10818223}, and accelerator availability~\cite{10.1145/3773276.3774299}. Orchestration must therefore ensure that placement and scaling decisions remain feasible under these constraints.
\vspace{-.4cm}
\subsubsection{\textbf{BA4: Elasticity implies available headroom.}} 
Serverless systems assume that additional capacity can be activated on demand.
\begin{itemize}[leftmargin=*, itemsep=1pt, topsep=2pt]
 \item \textbf{Breakage.} Execution is bounded by energy, thermal limits~\cite{xing2024deciphering}, and memory headroom~\cite{10818223}. Scaling decisions may violate feasibility even when computational resources are available~\cite{10818223}.
 \item \textbf{Implication.} Scaling must be \emph{feasibility-aware}, constrained by resource limits rather than driven solely by demand.
\end{itemize}
\vspace{-.2cm}
\subsubsection{\textbf{BA5: Execution configuration is static.}} 
Execution modes (e.g., CPU vs. GPU) are typically fixed or developer-defined.
\begin{itemize}[leftmargin=*, itemsep=1pt, topsep=2pt]
 \item \textbf{Breakage.} Static configurations fail under dynamic load and heterogeneous resources, leading to inefficient utilization or SLO violations~\cite{10.1145/3773276.3774299}.
 \item \textbf{Implication.} Execution must be \emph{adaptively configured}, with orchestration co-deciding placement and execution mode at runtime.
\end{itemize}

These breakage points reveal that execution feasibility is bounded by hard system constraints rather than elastic resource provisioning. This motivates the second research question (\textbf{RQ2}): ``How should an orchestrator control scaling and choose execution configurations so that workflow stages remain feasible under energy, onboard thermal, memory, and accelerator constraints?'' An orchestrator must therefore treat resource feasibility as a primary decision constraint, which leads to the following design implications: 
\begin{itemize}[leftmargin=*, itemsep=1pt, topsep=2pt]
\item \textbf{Energy- and thermal-aware admission control:} Admit or defer execution based on predicted resource margins.
\item \textbf{Time-aware execution planning:} Align execution with periods of sufficient resource availability.
\item \textbf{Dynamic execution-mode selection:} Adapt CPU/GPU configurations based on workload and system state.
\item \textbf{Accelerator-aware placement:} Allocate scarce heterogeneous resources efficiently under constraints.
\end{itemize}
\vspace{-.2cm}
\subsection{C3: Correctness and Progress under Decentralized and Delayed State} 
In the ECS continuum, the orchestrator must maintain forward progress despite intermittent connectivity, stale system state, and cross-domain constraints. Execution correctness can no longer rely on a single control plane and must instead be maintained under weak connectivity. 
\subsubsection{\textbf{BA6: Remote state is \textit{close enough}.}} 
The stateless-by-default serverless model externalizes state to managed services and typically assumes those services are reachable with predictable overhead. 
\begin{itemize}[leftmargin=*, itemsep=1pt, topsep=2pt]
 \item \textbf{Breakage.} Remote state access can dominate workflow latency and bandwidth consumption under intermittent connectivity and multi-hop paths~\cite{marcelino2025databelt}. This cost can become asymmetric when constrained uplinks limit state propagation and the exchange of intermediate data. 
 \item \textbf{Implication.} State must be \emph{explicitly placed and propagated}, treating data movement as rate-limited and direction-dependent rather than implicitly available.
\end{itemize}
\vspace{-.2cm}
\begin{table*}[t]
\centering
\caption{Mapping between broken assumptions, research questions, and orchestration implications.}
\label{tab:ba-rq-mapping}
\renewcommand{\arraystretch}{1.45}
\fontsize{7.5pt}{7.5pt}\selectfont
\begin{tabular}{p{0.6cm} p{4.9cm} p{1.6cm} p{6.4cm}}
\hline
\textbf{BA} & \textbf{Broken assumption} & \textbf{Mapped RQ} & \textbf{Implication for orchestration} \\
\hline

BA1 & Instantaneous reachability implies execution feasibility 
& RQ1 
& Execution must be temporally feasible over the stage lifetime, aligning placement and data exchange with predicted contact windows. \\

BA2 & Connectivity implies path feasibility 
& RQ1 
& Orchestration must ensure time-varying end-to-end path feasibility, accounting for dynamic topology and link characteristics. \\

BA3 & Execution binding is stable 
& RQ1 
& Execution must be time-indexed and re-bindable, supporting reassignment under mobility, handovers, and changing resource conditions. \\

BA4 & Elastic scaling assumes available resource headroom 
& RQ2 
& Scaling must be feasibility-aware, constrained by energy, thermal, and memory limits rather than reactive demand. \\

BA5 & Execution configuration is static 
& RQ2 
& The orchestrator must co-decide placement and execution mode, adapting configurations (e.g., CPU/GPU) at runtime. \\

BA6 & Remote state is \textit{close enough} 
& RQ3 
& State placement and propagation must be explicit decisions, treating data movement as rate-limited and asymmetric. \\

BA7 & Workflow handoff overhead is negligible 
& RQ3 
& Composition must be optimized via fusion and co-location to reduce data movement overheads. \\

BA8 & Cloud-style failure and recovery assumptions 
& RQ3 
& Recovery must be partition-tolerant, supporting checkpoint-aware execution under constrained coordination. \\

BA9 & Centralized control assumes timely and consistent global state 
& RQ3 
& Orchestration must tolerate stale and partial information, requiring prediction-aware decisions and delayed reconciliation. \\

BA10 & Single-domain uniform policy and governance 
& RQ3 
& Orchestration must be policy-aware, enforcing trust, compliance, and data movement constraints across domains. \\

\hline
\end{tabular}
\end{table*}
\subsubsection{\textbf{BA7: Workflow handoff overhead is negligible.}} 
Workflow composition of serverless functions assumes that inter-function data exchange is insignificant~\cite{li2022serverless}.
\begin{itemize}[leftmargin=*, itemsep=1pt, topsep=2pt]
 \item \textbf{Breakage.} Repeated state reads/writes and handoffs amplify latency and bandwidth overheads across constrained space-ground, inter-satellite, and edge-cloud paths~\cite{marcelino2025databelt}.
 \item \textbf{Implication.} Execution must be \emph{composition-aware}, applying fusion and co-location to reduce unnecessary data movement.
\end{itemize}
\vspace{-.2cm}
\subsubsection{\textbf{BA8: Cloud-style failure and recovery assumptions.}} 
Serverless systems assume frequent monitoring and rapid, centrally coordinated recovery.
\begin{itemize}[leftmargin=*, itemsep=1pt, topsep=2pt]
 \item \textbf{Breakage.} Recovery must operate under partitions and constrained coordination, where connectivity to control planes or replicas may be unavailable~\cite{xue2025space}.
 \item \textbf{Implication.} Execution must be \emph{partition-tolerant}, supporting checkpoint-aware scheduling~\cite{pfandzelter2024komet} and recovery mechanisms that remain feasible under intermittent connectivity.
\end{itemize}
\vspace{-.2cm}
\subsubsection{\textbf{BA9: Centralized control assumes timely and consistent state.}} 
Serverless orchestration typically relies on a centralized (or logically centralized) control plane operating on up-to-date global system state~\cite{li2022serverless}.
\begin{itemize}[leftmargin=*, itemsep=1pt, topsep=2pt]
 \item \textbf{Breakage.} In the ECS continuum, telemetry is delayed, partial, and dependent on intermittent connectivity~\cite{10818223}. The control plane operates on stale and potentially inconsistent views of the system~\cite{10818223}.
 \item \textbf{Implication.} Centralized orchestration must be \emph{robust to stale and partial information}, incorporating prediction, uncertainty-aware decision-making, and delayed state reconciliation.
\end{itemize}
\vspace{-.2cm}
\subsubsection{\textbf{BA10: Single-domain uniform policy and governance.}} 
Serverless systems assume a single administrative domain with uniform access and compliance.
\begin{itemize}[leftmargin=*, itemsep=1pt, topsep=2pt]
 \item \textbf{Breakage.} Execution spans multiple domains with heterogeneous governance, trust, and security constraints~\cite{10818223}, where data movement and execution must comply with domain-specific policies~\cite{shi2023comedge}. In orbital environments, orchestration may operate under adversarial conditions, including jamming, spoofed connectivity predictions, compromised nodes, or untrusted supply chains~\cite{xue2025space}.
 \item \textbf{Implication.} Orchestration must be \emph{policy- and trust-aware}, enforcing placement, data movement, and execution constraints as part of scheduling decisions while tolerating potentially unreliable or adversarial system information.
\end{itemize}
\begin{figure*}[!t]
	\centering
\includegraphics[width=.95\linewidth]{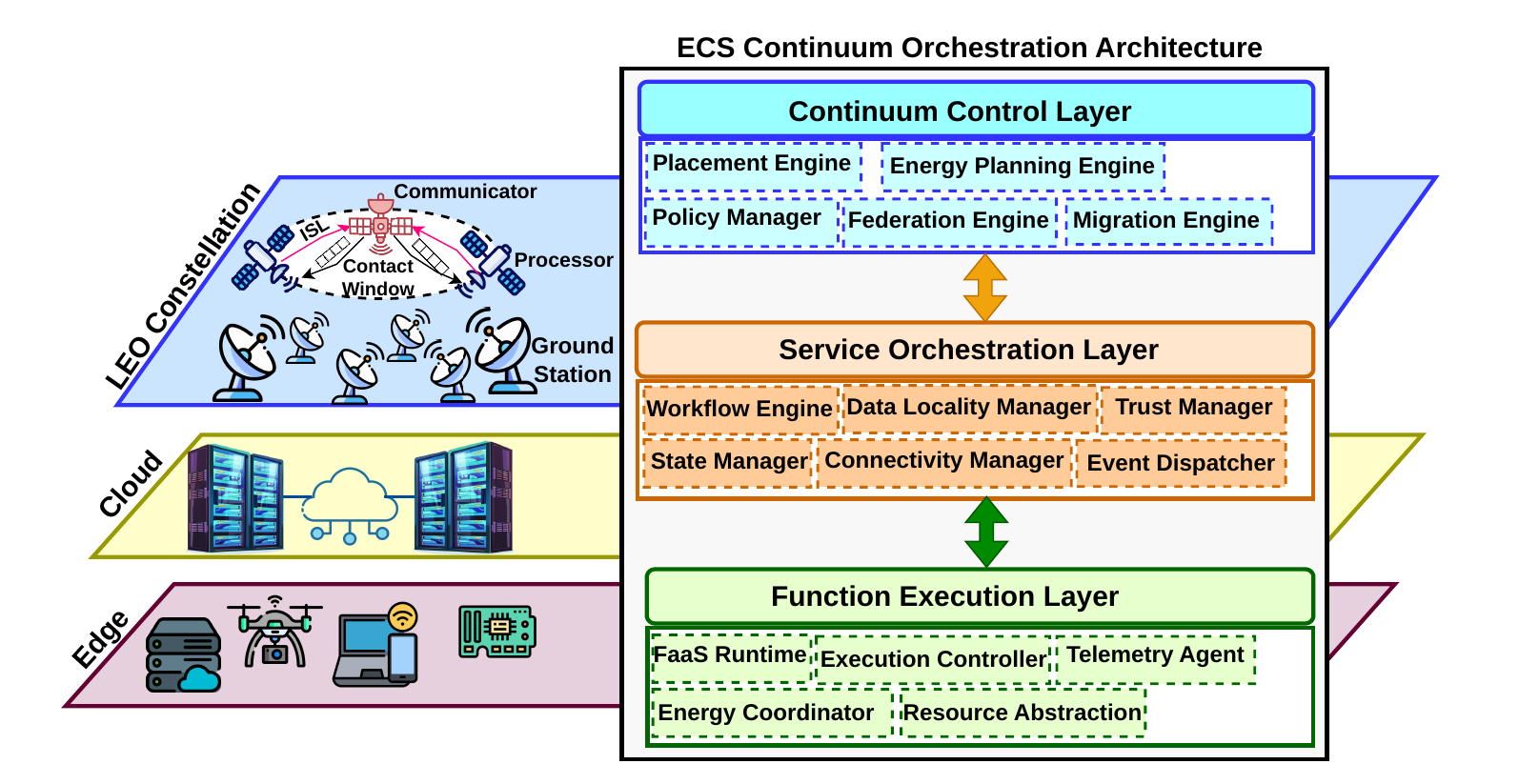}
	\caption{ECS continuum orchestration system architecture.}
	\label{arch}
\end{figure*}

These assumption failures show that correctness must be ensured under partial knowledge, constrained coordination, and cross-domain restrictions. The resulting research question (\textbf{RQ3}) is: ``How can an orchestrator ensure correctness and forward progress by managing state, composition, recovery, and policy under intermittent connectivity and stale system state?''
Addressing this requires the orchestrator to enforce progress-with-assurance under weak connectivity, leading to the following design implications:
\begin{itemize}[leftmargin=*, itemsep=1pt, topsep=2pt]
\item \textbf{State-aware placement and propagation:} Decide where the intermediate state resides and how it is staged or replicated to minimize remote access.
\item \textbf{Composition-aware optimization:} Apply fusion and co-location to reduce data movement overheads.
\item \textbf{Partition-tolerant progress and recovery:} Support checkpointing, rollback, and decentralized recovery under limited coordination.
\item \textbf{Decentralized control under stale telemetry:} Enable local autonomy with eventual consistency.
\item \textbf{Policy-aware orchestration across domains:} Enforce governance constraints on placement and data movement.
\end{itemize}

\section{ECS Continuum Orchestration Architecture}
\label{sec:sec3}
Fig.~\ref{arch} presents the proposed Edge-Cloud-Space (ECS) orchestration architecture, designed to address the fundamental mismatches identified in the previous section. The architecture explicitly models the continuum as a composition of edge nodes, cloud infrastructure, and LEO constellations. It assumes that a subset of satellites are \emph{processing}-capable (processors) and may expose heterogeneous compute profiles (e.g., CPU-only or accelerator-enabled nodes), while others primarily operate as \emph{communication relays} (communicators)~\cite{11316138}. These entities are interconnected via inter-satellite links (ISLs) and time-bounded ground contacts (\emph{contact windows}). Control decisions may be coordinated from terrestrial infrastructure, orbital nodes, or hybrid deployments, depending on connectivity and mission requirements. Consequently, orchestration operates under delayed, partial, and time-varying system visibility rather than assuming a continuously synchronized global state.

Unlike cloud-based designs, execution feasibility in this continuum is governed by time-varying connectivity, constrained resources, and delayed system state. The architecture, thus, separates concerns into three layers: \emph{(i)} a \emph{continuum control layer} for global, constraint-aware decision making, \emph{(ii)} a \emph{service orchestration layer} for workflow coordination under dynamic conditions, and \emph{(iii)} a \emph{function execution layer} for on-node execution of serverless functions. This decomposition directly aligns with the three challenges C1 (spatiotemporal feasibility), C2 (resource feasibility), and C3 (correctness under delayed and decentralized state).
\vspace{-0.2cm}

\subsubsection{\textbf{Revisiting the flood response workflow.}}
In the motivating scenario, initial flood detection is triggered at the edge, refined using satellite imagery, and stabilized in the cloud. This requires \emph{(i)} aligning execution with satellite contact windows (C1), \emph{(ii)} respecting onboard energy and compute limits (C2), and \emph{(iii)} ensuring progress despite delayed observations and cross-domain constraints (C3). The proposed architecture operationalizes these requirements as follows.
\vspace{-0.2cm}
\subsection{Continuum control layer}
This layer performs global, constraint-aware optimization over a logically centralized but physically distributed control plane. It explicitly addresses spatiotemporal feasibility (RQ1) and resource feasibility (RQ2).
\begin{itemize}[leftmargin=*, itemsep=1pt, topsep=2pt]
\item\textbf{Placement engine:}
Computes the initial placement of workflow stages by jointly considering node capabilities, predicted contact windows, and time-varying path feasibility~\cite{zhang2024utility}. This directly resolves BA1--BA3 by ensuring that placement decisions remain valid over the execution horizon rather than at invocation time.
\item\textbf{Energy planning engine.}
Models energy availability, thermal limits, and memory headroom across nodes, particularly in LEO processors. It enforces feasibility-aware scaling (BA4) and prevents invalid deployments under constrained onboard resources (RQ2).
\item\textbf{Migration engine:}
Enables time-indexed re-binding of execution by relocating functions when connectivity, resource availability, or associations change. It operationalizes continuity under mobility (BA3) and maintains feasibility across evolving contact windows (C1).
\item\textbf{Policy manager:}
Encodes SLOs, placement restrictions, and governance rules as hard constraints. It ensures that orchestration decisions remain policy-compliant across administrative domains (BA10) and integrates governance into scheduling. 
\item\textbf{Federation engine:}
Coordinates execution across multiple administrative domains (edge, cloud, satellite), enabling cross-domain orchestration while preserving trust and policy constraints, directly addressing BA10 and RQ3.
\end{itemize}
\vspace{-.5cm}
\subsection{Service orchestration layer}
This layer translates global decisions into executable workflows while maintaining consistency under intermittent connectivity and delayed state.
\begin{itemize}[leftmargin=*, itemsep=1pt, topsep=2pt]
\item\textbf{Workflow engine.}
Manages DAG-based workflow execution and dependency resolution under time-varying feasibility. It integrates scheduling decisions from the control layer and ensures stage execution aligns with contact opportunities (C1).
\item\textbf{Connectivity manager.}
Maintains a time-indexed view of the space-ground graph, including ISLs and contact windows. It enables spatiotemporal reasoning over links and paths, directly addressing BA1 and BA2 by exposing when and how communication is feasible.
\item\textbf{State manager.}
Maintains distributed workflow state and supports checkpointing and recovery. It enables partition-tolerant execution (BA8) and ensures forward progress under intermittent connectivity (C3).
\item\textbf{Data locality manager.}
Controls placement and movement of intermediate data across tiers, minimizing remote state access penalties (BA6) via co-locating computation and data.
\item\textbf{Trust manager.}
Ensures that sensitive operations remain within trusted domains and validates orchestration constraints under federated and potentially adversarial environments, directly addressing BA10.
\item\textbf{Event dispatcher.}
Implements event-driven invocation across the continuum. It transforms bursty inputs (e.g., flood reports) into coordinated workflow triggers and routes them under connectivity constraints.
\end{itemize}
\vspace{-.5cm}
\subsection{Function execution layer}
This layer realizes execution on edge devices, cloud servers, and LEO processors.
\begin{itemize}[leftmargin=*, itemsep=1pt, topsep=2pt]
\item\textbf{Function-as-a-Service (FaaS) runtime.}
Executes fine-grained functions with lightweight isolation across heterogeneous hardware.
\item\textbf{Execution controller.}
Implements local scheduling and enforces resource allocations under constraints imposed by upper layers.
\item\textbf{Resource abstraction.}
Provides a unified interface to heterogeneous compute resources and accelerator profiles (CPU, GPU, NPUs, domain-specific accelerators), enabling adaptive execution modes under heterogeneous onboard capabilities (BA5).
\item\textbf{Energy coordinator.}
Applies platform-exposed runtime power-management actions, such as throttling or concurrency limiting, where available, to enforce energy-aware decisions from the control layer (BA4).
\item\textbf{Telemetry agent.}
Continuously collects system state, including resource usage, energy, and connectivity. Given delayed and partial visibility (BA9), telemetry is used for prediction-aware and uncertainty-aware decision-making rather than instantaneous control.
\end{itemize}
\vspace{-.5cm}
\subsection{Cross-layer interaction and guarantees.}
The architecture forms a closed loop: telemetry flows upward from execution nodes, while placement, scaling, and migration decisions flow downward. Unlike cloud systems, this loop operates under a delayed and partial state, requiring predictive and time-indexed reasoning.
Across layers, the architecture enforces three invariants:
\begin{itemize}[leftmargin=*, itemsep=1pt, topsep=2pt]
\item\textbf{Spatiotemporal feasibility:} Execution and data movement are aligned with contact windows and time-varying paths.
\item\textbf{Resource feasibility:} Placement and scaling respect energy, thermal, and hardware constraints.
\item\textbf{Progress-with-assurance:} Workflows continue execution under partitions, delayed state, and cross-domain policies.
\end{itemize}

By explicitly integrating connectivity dynamics, resource constraints, and distributed control into orchestration decisions, the architecture addresses the ten broken assumptions (BA1--BA10) and provides a concrete blueprint for serverless execution across the edge--cloud--space continuum.

\section{ECS Orchestration Implementation Challenges}
\label{sec:sec4} 

The ECS architecture defines the orchestration responsibilities needed to execute serverless workflows across edge, cloud, and LEO infrastructures. However, realizing this architecture as an interoperable platform requires more than adding satellite nodes to existing serverless systems. The main gap is standardization: current FaaS platforms, workflow engines, telemetry stacks, and resource managers do not expose common abstractions for contact windows, time-varying paths, onboard feasibility limits, stale telemetry, or cross-domain policies. This section summarizes seven implementation challenges (\emph{IC1-IC7}) that must be addressed before ECS orchestration becomes a standardized platform capability.
\vspace{-.3cm}
\subsubsection{\textbf{IC1: Contact-window and path abstraction.}}
Existing schemes typically expose node availability and network metrics, but not time-indexed contact windows, ISL links, or predicted path feasibility. Standardized interfaces are needed for representing when links exist, how long they persist, and whether a workflow stage can exchange its required data within an SLO. This directly supports C1/RQ1 and the connectivity manager, placement engine, and workflow engine.
\vspace{-.2cm}
\subsubsection{\textbf{IC2: Feasibility-aware resource model.}}
Current serverless runtimes mostly expose CPU, memory, and concurrency limits. ECS execution requires richer models that capture energy budgets, onboard thermal state, accelerator availability, memory capacity, and communication cost~\cite{wang2025emulating}. Without a common feasibility model, placement and scaling remain cloud-style decisions that may be invalid on LEO processors or constrained edge nodes~\cite{10909133}. This supports C2/RQ2 and the energy planning engine, resource abstraction, execution controller, and energy coordinator.
\vspace{-.3cm}
\subsubsection{\textbf{IC3: Time-indexed orchestration APIs.}}
Standard FaaS and workflow systems are invocation-centric. They trigger functions in response to events and scale based on demand signals such as request rate, queue depth, or concurrency. ECS workflows require plans over execution horizons. APIs must support deferred execution, validity intervals, contact-aware reservations, re-binding, and migration triggers. This connects C1 and C2 to the placement, migration, and workflow engines.
\vspace{-.3cm}
\subsubsection{\textbf{IC4: State and data locality primitives.}}
Serverless platforms externalize state to storage services and assume remote access overhead remains within acceptable bounds. ECS orchestration requires explicit state placement, staging, replication, checkpointing, and function-state fusion to avoid repeated multi-hop transfers~\cite{zengshan2025comprehensive}. These primitives are central to C3/RQ3 and map to the state and data locality managers.
\vspace{-.3cm}
\subsubsection{\textbf{IC5: Stale telemetry and uncertainty handling.}}
Cloud controllers assume fresh telemetry. In ECS, telemetry can be delayed, partial, or unavailable during partitions. Implementations, therefore, need telemetry freshness metadata, confidence intervals, prediction hooks, and reconciliation mechanisms. This supports C3/RQ3 and maps to the telemetry agent, connectivity manager, and control layer.
\vspace{-.3cm}
\subsubsection{\textbf{IC6: Cross-domain policy and trust interface.}}
ECS workflows cross administrative boundaries among edge operators, cloud providers, satellite operators, and data owners. Standard orchestration interfaces must express placement constraints, data locality rules, trust levels, identity, authorization, and compliance requirements. This supports BA10/C3 and maps to the trust manager, policy manager, and federation engine.
\vspace{-.3cm}
\subsubsection{\textbf{IC7: Failure semantics under intermittent connectivity.}}
Current platforms usually define retry, timeout, and failure semantics under the assumption of continuous control-plane reachability. ECS execution needs partition-aware semantics, i.e., when to retry, defer, checkpoint, migrate, or continue locally under disconnected operation. This supports C3/RQ3 and maps to the state manager, execution controller, and migration engine.
\begin{table*}[t]
\centering
\caption{Implementation challenges and their mapping to ECS orchestration requirements.}
\label{tab:implementation_challenges}
\renewcommand{\arraystretch}{1.3}
\fontsize{7.5pt}{7.5pt}\selectfont
\begin{adjustbox}{width=1\textwidth}
\begin{tabular}{p{0.5cm} p{4.1cm} p{2.4cm} p{3.4cm} p{4.8cm}}
\hline
\textbf{ID} & \textbf{Implementation challenge} & \textbf{Main RQ/C} & \textbf{Key components} & \textbf{Required standardization gap} \\
\hline
IC1 & Contact-window and path abstraction & RQ1 / C1 & Connectivity manager, placement engine, workflow engine & Common model for contact windows, ISLs, ground links, path validity, and transfer feasibility. \\
IC2 & Feasibility-aware resource model & RQ2 / C2 & Energy planning engine, resource abstraction, execution controller, energy coordinator & Standard resource descriptors for energy budget, thermal headroom, accelerator availability, memory pressure, and communication cost. \\
IC3 & Time-indexed orchestration APIs & RQ1--RQ2 / C1--C2 & Placement engine, migration engine, workflow engine & APIs for deferred execution, validity intervals, reservations, migration, and re-binding. \\
IC4 & State and data locality primitives & RQ3 / C3 & State manager, data locality manager & Portable primitives for state placement, staging, replication, checkpointing, and function-state fusion. \\
IC5 & Stale telemetry and uncertainty handling & RQ3 / C3 & Telemetry agent, connectivity manager, continuum control layer & Telemetry freshness, confidence metadata, prediction hooks, and reconciliation semantics. \\
IC6 & Cross-domain policy and trust interface & RQ3 / C3 & Trust manager, policy manager, federation engine & Common policy model for trust, identity, locality, authorization, and compliance. \\
IC7 & Failure semantics under intermittent connectivity & RQ3 / C3 & State manager, execution controller, migration engine & Standard semantics for retry, defer, checkpoint, migrate, and local execution under disconnection. \\
\hline
\end{tabular}
\end{adjustbox}
\end{table*}

Overall, these challenges show that ECS orchestration is not only an implementation problem but also a standardization problem. Existing platforms provide useful building blocks, but they lack shared abstractions for time, feasibility, state, and policy across space-ground environments. Closing this gap is necessary before serverless orchestration can become a portable execution model for the edge-cloud-space continuum.

\section{ECS Orchestration in Action: Representative Use Cases}
\label{sec:sec5}
To demonstrate how the proposed ECS architecture operates in practice, we revisit and extend the motivating \emph{flood response} scenario as a representative use case from the \emph{disaster response and humanitarian operations} domain. This scenario stresses all three orchestration challenges (C1--C3) and exposes the necessity of ECS-specific design decisions.
\vspace{-0.2cm}
\subsection{Use Case Overview: Flood Detection and Response}
A flood event is initially detected by edge sensors (e.g., river gauges, cameras). The objective is to generate timely situational awareness, including flood extent maps and evacuation recommendations, by combining edge data, satellite imagery, and cloud-based analytics. The workflow consists of three stages:
\begin{enumerate}
\item \textbf{Edge detection:} Lightweight anomaly detection is triggered at edge nodes.
\item \textbf{Satellite refinement:} High-resolution imagery is processed onboard LEO satellites during contact windows.
\item \textbf{Cloud aggregation:} Results are consolidated, fused with historical data, and disseminated to authorities.
\end{enumerate}

This workflow spans heterogeneous domains and must execute under intermittent connectivity, constrained resources, and delayed system state.
\vspace{-0.5cm}
\subsection{Mapping to ECS Architecture}
The execution of this workflow is realized through coordinated interaction across the three ECS layers (Fig.~\ref{arch}).
\vspace{-0.2cm}
\subsubsection{\textbf{Continuum control layer:}} Determines \emph{where and when} each stage executes:
\begin{itemize}[leftmargin=*, itemsep=1pt, topsep=2pt]
\item \textbf{Placement engine:}
Schedules the satellite refinement stage only within predicted \emph{contact windows}, ensuring that image acquisition, processing, and result transmission are temporally feasible~\cite{liu2025device} (addresses BA1--BA2).
\item \textbf{Energy planning engine:}
Selects satellites with sufficient onboard energy and thermal headroom to execute image processing tasks, avoiding infeasible deployments (addresses BA4).
\item \textbf{Migration engine:}
Reassigns processing to alternative satellites when contact conditions change or handovers occur, maintaining execution continuity (addresses BA3).
\item \textbf{Policy and federation engines:}
Ensure that sensitive data (e.g., critical infrastructure maps) is processed only within trusted domains (e.g., cloud or authorized satellites), enforcing cross-domain constraints (addresses BA10).
\end{itemize}
\vspace{-0.5cm}
\subsubsection{\textbf{Service orchestration layer:}} Ensures \emph{correct execution progression} under dynamic conditions:
\begin{itemize}[leftmargin=*, itemsep=1pt, topsep=2pt]
\item \textbf{Workflow engine:}
Coordinates progressive, dependency-aware workflow execution so that stages can run, defer, or resume as inputs, contact windows, and feasible data paths become available. 
\item \textbf{Connectivity manager:}
Maintains a time-indexed view of ISLs and ground contacts, enabling the system to decide whether intermediate data can be transferred within deadlines (addresses BA2).
\item \textbf{State manager:}
Stores intermediate flood maps and supports checkpointing when connectivity is lost, enabling continuation after partitions (addresses BA8).
\item \textbf{Data locality manager:}
Ensures that image processing is performed close to where data is generated (e.g., onboard satellites), minimizing costly multi-hop transfers (addresses BA6--BA7).
\item \textbf{Event dispatcher:}
Transforms sensor-triggered flood alerts into workflow executions across the continuum.
\end{itemize}
\vspace{-0.5cm}
\subsubsection{\textbf{Function execution layer:}} Realizes execution under node-level constraints:
\begin{itemize}[leftmargin=*, itemsep=1pt, topsep=2pt]
\item \textbf{FaaS runtime:}
Executes flood detection and image processing functions across edge, cloud, and LEO nodes.
\item \textbf{Execution controller:}
Enforces local scheduling decisions and resource limits, ensuring compliance with energy and compute constraints (addresses BA4--BA5).
\item \textbf{Resource abstraction:}
Enables adaptive execution modes (CPU vs. GPU) depending on available hardware.
\item \textbf{Telemetry agent:}
Continuously reports system state (e.g., energy, load, connectivity), although telemetry may be delayed, incomplete, or stale. The control layer uses this information for prediction-aware decisions (addresses BA9).
\end{itemize}
\vspace{-0.5cm}
\subsection{ECS Orchestration Invariants}
The key reason this architecture is suitable is that it enforces three invariants that directly address the broken assumptions:
\begin{itemize}[leftmargin=*, itemsep=1pt, topsep=2pt]
\item \textbf{Spatiotemporal feasibility:}
Execution is aligned with contact windows and time-varying paths, ensuring that all stages complete within connectivity constraints.
\item \textbf{Resource feasibility:}
Placement and execution respect energy, thermal, and hardware limits, avoiding invalid deployments.
\item \textbf{Progress-with-assurance:}
Workflow execution continues under partitions via checkpointing, state propagation, and policy-aware control.
\end{itemize}

In contrast to conventional serverless systems, which assume continuous connectivity, elastic resources, and fresh global state, ECS orchestration explicitly models \emph{time}, \emph{feasibility}, and \emph{uncertainty} as first-class concerns. This enables reliable execution of cross-domain workflows such as flood response, where correctness and timeliness depend on coordinated operation across edge, cloud, and space.
\vspace{-0.3cm}
\subsection{Generalization Across Domains}
Although demonstrated for disaster response, the same orchestration pattern applies across other domains (e.g., environmental monitoring, maritime surveillance, infrastructure inspection)~\cite{11036255}. In all cases, workflows exhibit time-dependent connectivity, constrained execution resources, and cross-domain coordination requirements. Thus, the ECS architecture provides a general blueprint for serverless orchestration across the edge-cloud-space continuum.

\section{Conclusion}
\label{sec:sec6}
This article identified fundamental mismatches between conventional serverless orchestration and the requirements of the edge-cloud-space continuum, formalized as ten broken assumptions across three core challenges: spatiotemporal feasibility, resource feasibility, and progress-with-assurance under delayed and decentralized state. To address these gaps, we presented a multi-layer ECS orchestration architecture that elevates time-varying connectivity, hard resource constraints, and stale system state to first-class concerns, providing a concrete, implementable model aligned with the raised research questions. We further highlighted key implementation gaps that prevent current platforms from supporting ECS orchestration and demonstrated, through a representative flood-response workflow, how the architecture enables correct and feasible execution across heterogeneous and intermittently connected domains. The next step is to turn these architectural responsibilities into portable orchestration interfaces, so that serverless platforms can treat time, feasibility, and trust as first-class execution concerns in space-ground environments.

\bibliographystyle{IEEEtran}
\bibliography{bibtex}
{\small
\authorbio{Hadi Tabatabaee Malazi}
{is an Assistant Professor in the School of Computer Science at University College Dublin, Ireland. He leads the Sustainable Orchestration in Computing Continuum (SOC\textsuperscript{2}) Lab and is an IEEE Senior Member. His research focuses on sustainable and resilient services computing. He is particularly interested in carbon-aware, constraint-aware, and reliability-aware orchestration of distributed services across edge–cloud, multi-cloud, and space-enabled computing environments. Contact him at: hadi.tabatabaeemalazi@ucd.ie.}

\authorbio{Reza Farahani}
{(\textbf{Corresponding author}) is a Postdoctoral Researcher (University Assistant) at the Distributed Systems Group (DSG), TU Wien, Austria, and a Lecturer at the University of Klagenfurt, Austria. He has recently coordinated the Austrian EdgeAI-Drone project and participated in the ENFIELD Exchange Scheme funded by the European Union, focusing on Green AI. From 2019 to 2023, he was with the Christian Doppler Laboratory ATHENA and, from 2023 to 2026, contributed to the EU Graph-Massivizer project, where he led WP5 on serverless orchestration and large-scale edge–cloud testbeds. 
He was a Visiting Scholar at the University of Surrey, UK. 
His research interests include distributed and networked systems, edge-cloud continuum, serverless computing, distributed multimedia, agentic AI, and Green AI. Contact him at: r.farahani@dsg.tuwien.ac.at.}

\authorbio{Nitinder Mohan}
{is an Assistant Professor at the Delft University of Technology (TU Delft), Netherlands where he leads the Systems and Protocols for Edge-Enabled Internet (SPEAR) lab associated with Networked Systems group. His research has been awarded with several awards such as "Outstanding Ph.D. Dissertation Award 2020" by the IEEE TCSC, IETF/IRTF Applied Networking Research Prize, and several best paper awards at IEEE and ACM venues. His research interests are in edge computing, orchestration systems, next-generation cyber-physical systems, satellite networks, and wide-scale Internet measurements. Contact him at n.mohan@tudelft.nl.}

\authorbio{Schahram Dustdar}
{is a Full Professor of Computer Science and Head of Distributed Systems Group (DSG) at TU Wien, Vienna, Austria, and is affiliated with ICREA, Barcelona, Spain. He holds several honorary positions, including Honorary Professor at Nanjing University of Science and Technology (since 2023) and Francqui Chair Professor at the University of Namur (2021–2022). He has held visiting positions at institutions including the University of Southern California (Los Angeles), Monash University, Shanghai University, Macquarie University, Universitat Pompeu Fabra (Barcelona), the University of Sevilla, and UC Berkeley. Contact him at dustdar@dsg.tuwien.ac.at.}
\end{document}